\documentclass[prd, letter, preprintnumbers, amsmath, amssymb, nofootinbib, aps, 12pt]{revtex4-2}

\usepackage{bm}
\usepackage{amsfonts}
\usepackage{mathrsfs}
\usepackage{graphicx}
\usepackage{amsmath}
\usepackage{float}
\usepackage{braket}
\usepackage[paperwidth=8.5in,paperheight=11in,centering,hmargin=2.5cm,vmargin=2.5cm]{geometry}

\begin{document}

\newcommand{\sgn}{\operatorname{sgn}}
\newcommand{\hhat}[1]{\hat {\hat{#1}}}
\newcommand{\pslash}[1]{#1\llap{\sl/}}
\newcommand{\kslash}[1]{\rlap{\sl/}#1}
\newcommand{\lab}[1]{}
\newcommand{\iref}[2]{}
\newcommand{\sto}[1]{\begin{center} \textit{#1} \end{center}}
\newcommand{\rf}[1]{{\color{blue}[\textit{#1}]}}
\newcommand{\eml}[1]{#1}

\newcommand{\er}[1]{Eq.\eqref{#1}}
\newcommand{\df}[1]{\textbf{#1}}
\newcommand{\mdf}[1]{\pmb{#1}}
\newcommand{\ft}[1]{\footnote{#1}}
\newcommand{\n}[1]{$#1$}
\newcommand{\fals}[1]{$^\times$ #1}
\newcommand{\new}{{\color{red}$^{NEW}$ }}
\newcommand{\ci}[1]{}
\newcommand{\de}[1]{{\color{green}\underline{#1}}}
\newcommand{\ke}{\rangle}
\newcommand{\br}{\langle}
\newcommand{\lb}{\left(}
\newcommand{\rb}{\right)}
\newcommand{\lbk}{\left[}
\newcommand{\rbk}{\right]}
\newcommand{\blb}{\Big(}
\newcommand{\brb}{\Big)}
\newcommand{\nn}{\nonumber \\}
\newcommand{\p}{\partial}
\newcommand{\pd}[1]{\frac {\partial} {\partial #1}}
\newcommand{\cd}{\nabla}
\newcommand{\cc}{$>$}
\newcommand{\bqa}{\begin{eqnarray}}
\newcommand{\eqa}{\end{eqnarray}}
\newcommand{\bqe}{\begin{equation}}
\newcommand{\eqe}{\end{equation}}
\newcommand{\bay}[1]{\left(\begin{array}{#1}}
\newcommand{\eay}{\end{array}\right)}
\newcommand{\eg}{\textit{e.g.} }
\newcommand{\ie}{\textit{i.e.}, }
\newcommand{\iv}[1]{{#1}^{-1}}
\newcommand{\st}[1]{|#1\ke}
\newcommand{\at}[1]{{\Big|}_{#1}}
\newcommand{\zt}[1]{\texttt{#1}}
\newcommand{\non}{\nonumber}
\newcommand{\m}{\mu}

\def\xa{{m}}
\def\xA{{m}}
\def\xb{{\beta}}
\def\xB{{\Beta}}
\def\xd{{\delta}}
\def\xD{{\Delta}}
\def\xe{{\epsilon}}
\def\xE{{\Epsilon}}
\def\xve{{\varepsilon}}
\def\xg{{\gamma}}
\def\xG{{\Gamma}}
\def\xk{{\kappa}}
\def\xK{{\Kappa}}
\def\xl{{\lambda}}
\def\xL{{\Lambda}}
\def\xo{{\omega}}
\def\xO{{\Omega}}
\def\xvp{{\varphi}}
\def\xs{{\sigma}}
\def\xS{{\Sigma}}
\def\xt{{\theta}}
\def\xvt{{\vartheta}}
\def\xT{{\Theta}}
\def \Tr {{\rm Tr}}
\def\CA{{\cal A}}
\def\CC{{\cal C}}
\def\CD{{\cal D}}
\def\CE{{\cal E}}
\def\CF{{\cal F}}
\def\CH{{\cal H}}
\def\CJ{{\cal J}}
\def\CK{{\cal K}}
\def\CL{{\cal L}}
\def\CM{{\cal M}}
\def\CN{{\cal N}}
\def\CO{{\cal O}}
\def\CP{{\cal P}}
\def\CQ{{\cal Q}}
\def\CR{{\cal R}}
\def\CS{{\cal S}}
\def\CT{{\cal T}}
\def\CV{{\cal V}}
\def\CW{{\cal W}}
\def\CY{{\cal Y}}
\def\BC{\mathbb{C}}
\def\BR{\mathbb{R}}
\def\BZ{\mathbb{Z}}
\def\sA{\mathscr{A}}
\def\sB{\mathscr{B}}
\def\sF{\mathscr{F}}
\def\sG{\mathscr{G}}
\def\sH{\mathscr{H}}
\def\sJ{\mathscr{J}}
\def\sL{\mathscr{L}}
\def\sM{\mathscr{M}}
\def\sN{\mathscr{N}}
\def\sO{\mathscr{O}}
\def\sP{\mathscr{P}}
\def\sR{\mathscr{R}}
\def\sQ{\mathscr{Q}}
\def\sS{\mathscr{S}}
\def\sX{\mathscr{X}}

\def\slz{SL(2,Z)}
\def\slr{$SL(2,R)\times SL(2,R)$ }
\def\ads{${AdS}_5\times {S}^5$ }
\def\adst{${AdS}_3$ }
\def\sun{SU(N)}
\def\ad#1#2{{\frac \delta {\delta\sigma^{#1}} (#2)}}
\def\bqf{\bar Q_{\bar f}}
\def\nf{N_f}
\def\sunf{SU(N_f)}

\def\dcirc{{^\circ_\circ}}

\author{Morgan Henry Lynch}
\email{corresponding author - morgan.lynch@technion.ac.il}
\affiliation{Department of Electrical Engineering,
Technion: Israel Institute of Technology,\\ Haifa 32000, Israel}

\title{Notes on the experimental observation of the Unruh effect}
\date{\today}

\begin{abstract}
The incorporation of classical general relativity into quantum field theory yields a surprising result - thermodynamic particle production. One such phenomenon, known as the Unruh effect, causes empty space to effervesce a thermal bath of particles when viewed by an observer undergoing uniformly accelerated motion. These systems will have a Rindler horizon which produces this Unruh radiation at the Fulling-Davies-Unruh temperature. For accelerated charges, the emission and absorption of this radiation will imprint the FDU temperature on photons emitted in the laboratory. Each of these photons will also change the Rindler horizon in accordance with the Bekenstein-Hawking area-entropy law. In this essay, we will discuss these aspects of acceleration-induced thermality which have been experimentally observed in a high energy channeling experiment carried out by CERN-NA63. \\ \; \\ \; \\
\begin{center} \textbf{Essay written for the Gravity Research Foundation \\2022 Awards for Essays \\on Gravitation.} \end{center}
\end{abstract}


\maketitle
\clearpage
\section{Thermodynamic Particle Production}

One of the key predictions of quantum field theory in curved spacetime \cite{davies, aspects, pt, parker,parker1} is the creation of thermalized particles from the vacuum. A wide variety of techniques have been implemented to analyze this phenomenon including the response of a two level system, known as an Unruh-DeWitt detector \cite{unruh1, dewitt}, when it absorbs these particles. One case, in particular, is that of acceleration in vacuum. There, an Unruh-DeWitt detector will absorb one of these thermal particles, undergo a transition, either up or down in energy, and radiate \cite{lynch,lynch1,muller,matsas1,matsas3}. Remarkably, the character of the radiation emitted by the detector will also be thermalized at a temperature defined by the acceleration, $T = \frac{a}{2 \pi}$, known as the Fulling-Davies-Unruh (FDU) temperature. The path towards the discovery of this FDU temperature began with the analysis of the vacuum structure of both inertial and accelerated reference frames by Fulling \cite{fulling}, the radiation flux from a 1+1 dimensional moving mirror by Davies \cite{davies1}, and finally by analyzing Hawking radiation in the near horizon limit of black holes by Unruh \cite{unruh1}. 

These pursuits culminated in an inherently thermodynamic understanding of quantum field theories in classical general relativistic backgrounds \cite{jacobson}. One of key tenets of this thermodynamic interpretation comes from the entropy of black holes. The work of Bekenstein \cite{bekenstein} and Hawking \cite{hawking} demonstrated that the entropy, $S$, associated with a blackhole would be proportional to the surface area, $A$, of the event horizon; namely $S = \frac{A}{4}$. This became known as the Bekenstein-Hawking area-entropy (BHAE) law. This area-entropy law is also applicable to the Rindler horizon associated with acceleration \cite{satz}. Consequently, given a system with sufficient acceleration, thermality can be verified and explored by both the presence of a well-defined FDU temperature as well as the BHAE law of the Rindler horizon.

Since the FDU temperature, Rindler horizon dynamics, and the general characteristics of electromagnetic radiation are determined by acceleration, it is through accelerated electromagnetic systems that we expect our first insights into these phenomena. The high energy channeling radiation experiments carried out by CERN-NA63 \cite{wistisen} have provided such a scenario. These experiments, with a high energy positron penetrating a sample of single crystal, were successful in measuring radiation reaction. The immense recoil produced by the radiation reaction also provided us with the enormous accelerations necessary to experimentally observe these intriguing aspects of acceleration-induced thermality \cite{aqd}.

\section{Thermalized power spectrum}

Uniformly accelerated charges will radiate photons in accordance with the Larmor formula. At the same time, for sufficiently large accelerations, one would expect recoil due to the photon emission as well as the radiation emitted to reflect a thermalization due to the Unruh effect. As such, we will incorporate these two phenomena into the process of photon emission and compare it to a high energy channeling radiation experiment carried out by CERN-NA63. If we consider a semiclassical vector current coupled to an Unruh-DeWitt detector to model an accelerated charge, we will have the following current, $\hat{j}_{\m}(x) = qu_{\m}\hat{m}(\tau)\delta^{3}(x-x_{tr}(\tau))$. Here, $u_{\m}$ is the four-velocity, $\hat{m}(\tau)$ is the monopole moment operator, $q$ is the charge, and $x_{tr}(\tau)$ is the trajectory of the emitter. Then, to analyze a radiative photon emission process using the QED current interaction, $\hat{S}_{I} = \int d^{4}x \hat{j}_{\m}(x)\hat{A}^{\m}(x)$, we arrive at the response function, or photon emission rate, of the Unruh-DeWitt detector \cite{lynch4},
\bqa
\Gamma = q^{2} \int d\xi e^{-i\Delta E \xi}  u_{\m}u_{\nu}[x',x]G^{\m \nu}[x',x].
\label{response}
\eqa

We find the standard Fourier transform of the vacuum-to-vacuum photon Wightman function, $G^{\nu \m}[x',x] = \bra{0} \hat{A}^{\dagger \nu}(x')\hat{A}^{\m}(x)\ket{0}$. We wish to compute the power spectrum, $\frac{d\mathcal{S}}{d\omega} = \frac{d \lb \Gamma \omega \rb}{d \omega}$, from the above response function. Our trajectory will have a constant proper acceleration $a$, parametrized by the proper time $\tau$, with a four-velocity $u_{\m} = (\cosh{(a\tau)},0,0,\sinh{(a\tau)})$. Thus, we find \cite{aqd}, 
\bqa
\frac{d \mathcal{S}}{d \omega} &=& -i \frac{4}{3}\alpha \frac{ \omega^2}{a} \lbk  \delta H^{(2)}_{\frac{2i\Delta E}{a}}\lb - \frac{2i \omega \gamma}{a} \rb  -\frac{1}{2} \lb  H^{(2)}_{\frac{2i\Delta E}{a} -2}\lb - \frac{2i\omega  \gamma}{a} \rb +  H^{(2)}_{\frac{2i\Delta E}{a} +2}\lb - \frac{2i\omega  \gamma}{a} \rb \rb \rbk \non \\
&\times & 2\sinh{(\pi \Delta E/a)}  \coth{(\Delta E/(2T_{r}))} e^{\frac{\pi \Delta E}{a}}. \label{temp}
\eqa
Here, we have defined the rapidity parameter $\delta = 2 \gamma^{2}-1$. Note the presence of thermality manifests via the Hankel identity, $H^{(2)}_{\frac{2i\Delta E}{a}}(x) = e^{-\Delta E/T_{FDU}}H^{(2)}_{-\frac{2i\Delta E}{a}}(x)$, which reflects the detailed balance at thermal equilibrium. Also note we have the thermal factor which explicitly encodes the emission and absorption of Rindler photons from the Rindler bath at an arbitrary temperature, $T_{r}$. This expression reduces to the statement of detailed balance upon substitution of $T_{r} = T_{FDU}$ and $\omega_{r} = \Delta E$ \cite{matsas5}. Hence, 
\bqe
 2\sinh{(\pi \omega_{r}/a)}  \coth{(\omega_{r}/(2T_{r}))} e^{\frac{\pi \omega_{r}}{a}} \rightarrow 
\lbk 1+ e^{\Delta E/T_{FDU}} \rbk.\label{conv}
\eqe
What the above expression illustrates is the \textit{detailed balance of the Unruh-DeWitt detector} $\leftrightarrow$ \textit{photon exchange with Rindler bath in Rindler space}. Our energy gap is given by, $\Delta E \sim \Omega + \Omega A \omega +\frac{\omega^{2}}{2m}$; the sum of a pure channeling frequency $\Omega$ \cite{frolov}, the Rindler frequency $\omega_{r} = A\Omega \omega$ \cite{matsas5, kolekar}, and the recoil kinetic energy term $\frac{\omega^{2}}{2m}$ \cite{lynch4, sokolov}.

The acceleration of these systems comes from the radiation reaction itself. If we consider a characteristic lab frame photon frequency, $\omega_{0}$, produced by the channeling radiation, we find our proper acceleration to be, $a' = \frac{\omega_{0}^{2}}{ \pi m}$. The associated temperature is then given by,
\bqe
T_{RR} = \frac{\omega^{2}_{0}}{ 2 m \pi^{2} }.
\eqe

The acceleration/temperature scale of the system is set by the recoil energy of the first frequency to thermalize, in this case $\omega_{0} \sim 150$ GeV. This immense recoil acceleration, $a \sim 6 \times 10^{39}$ $m/s^{2}$, gives rise to a temperature of $2 \times 10^{19}$ K or about $2$ PeV. The incredibly high energy scales here are due specifically to the presence of a robust radiation reaction which imparts a fully relativistic recoil momentum onto the positron during the emission time of a single photon period. 

The thermalization time is also given by, $t(\omega) = \frac{1}{\int_{0}^{\omega}\frac{d\mathcal{S}(\omega')}{d\omega'} \frac{1}{\omega'}d\omega'}$. Thermality necessitates that the emission lifetime is shorter than the experimental lifetime. This implies a low frequency cutoff for frequencies that do not meet this criterion. The reduced chi-squared statistic for each low frequency cutoff is then computed. The thermalization time and chi-squared values for each low frequency cutoff are presented in Fig. 1 below.

\begin{figure}[H]
\centering  
\includegraphics[scale=.35]{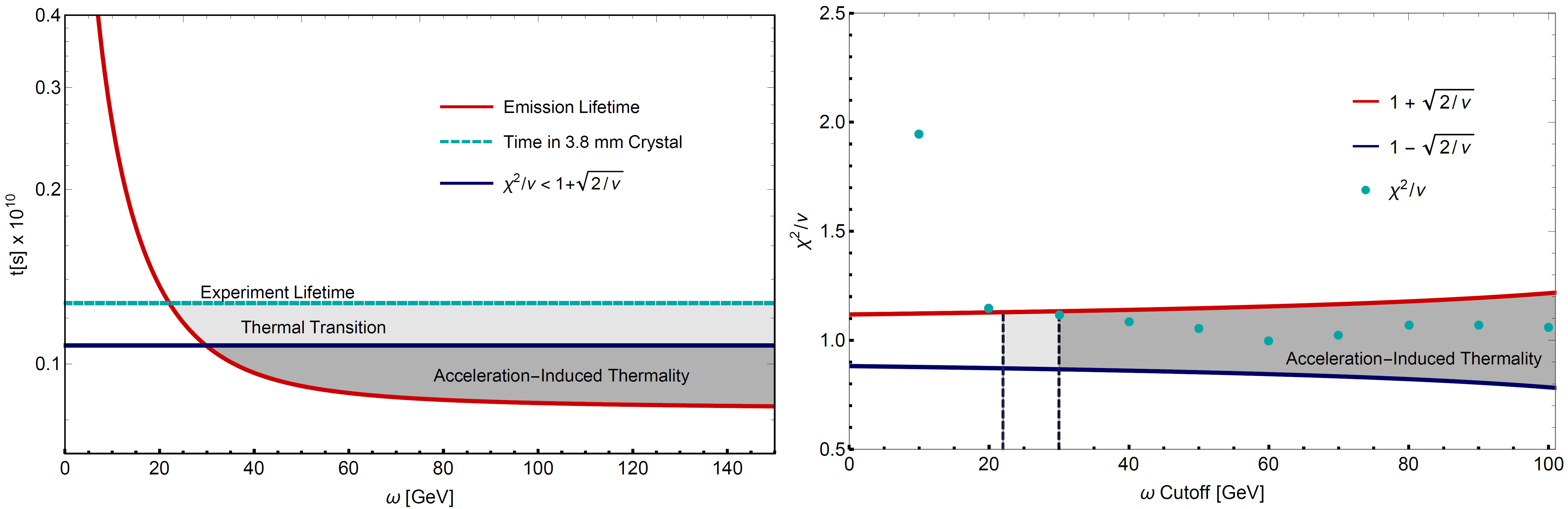} 
  \caption{\textbf{Thermalization time and chi-squared statistic:} (Left) demonstrates the system has time to thermalize beyond the threshold energy of $\sim 22$ GeV. (Right) shows the chi-squared statistic for each best fit power spectrum with low frequency cutoffs at multiples of $10 \;$ GeV. }
\end{figure}

The power spectrum, Eq. (2), for the first cutoff, $30 \;$ GeV, with a chi-squared statistic below the one standard deviation threshold is presented in Fig. 2 below. By comparing this power spectrum to the channeling radiation data set, we can find best fits for acceleration in order to measure $T_{FDU} = \frac{a}{2 \pi}$. Moreover, using the thermal conversion factor, Eq. (3), we can also directly measure the temperature of the Rindler bath $T_{r}$ from the data set. 

\begin{figure}[H]
\centering  
\includegraphics[scale=.35]{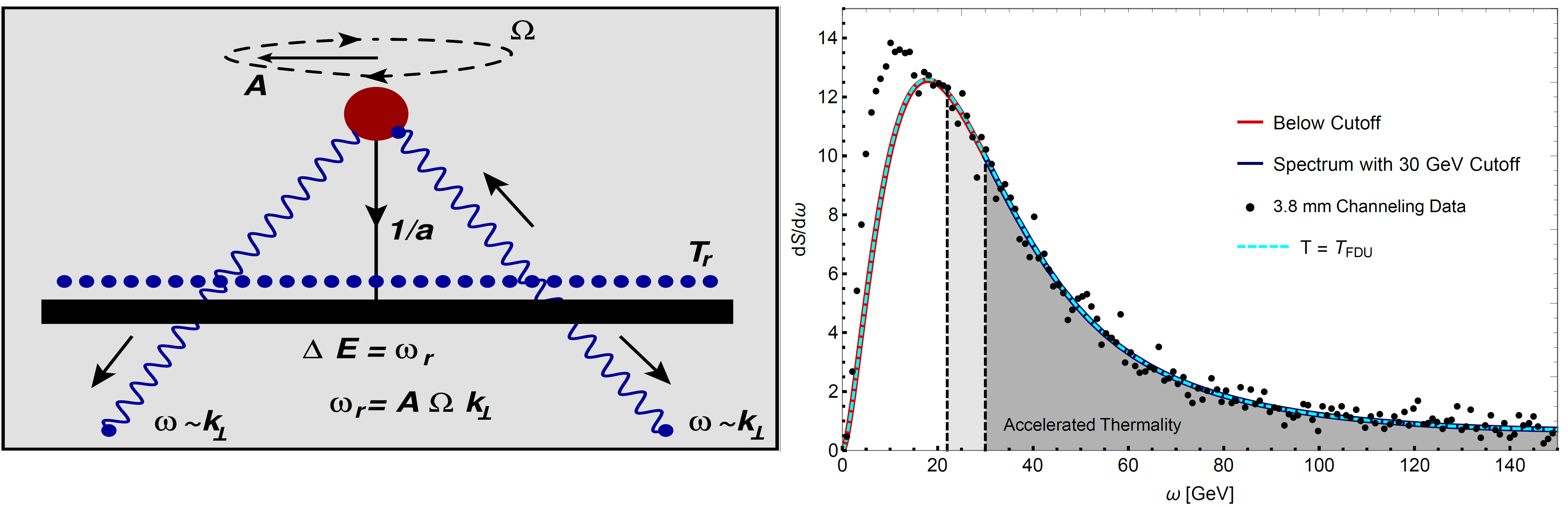} 
  \caption{\textbf{Rindler photon exchange and thermal power spectrum:} (Left) demonstrates the photon exchange with the Rindler bath at the characteristic Rindler energy set by the transverse oscillation \cite{matsas5, kolekar}. (Right) shows the power spectrum for the $30$ GeV cutoff, the first cutoff to yield a $\chi^{2}/\nu$ below the 1 standard deviation threshold, and the best fit Rindler temperature equal to the Fulling-Davies-Unruh temperature.}
\end{figure}

The measurement of the accelerations give rise to $T_{FDU} = 1.80 \pm .51 \; PeV $ and the temperature of the Rindler bath is found to be $T_{R} = 1.96 \pm .49 \; PeV$. For comparison, the upper bound on the temperature produced by the recoil acceleration is given by $T_{RR} = 2.23 \; PeV$. As such, we find
\bqe
T_{R} = T_{FDU}(1.09 \pm .41)
\eqe
These experimental measurements confirm the predictions of Fulling \cite{fulling}, Davies \cite{davies1}, and Unruh \cite{unruh1}. This also realizes the proposal put forward by Cozzella et al. for directly measuring the FDU temperature  \cite{matsas5}.

\section{The Bekenstein-Hawking Area-Entropy Law}

Bekenstein reasoned that the entropy of a black hole would be proportional to the surface area. Hawking, upon the discovery of black hole evaporation, was able to fix the proportionality constant to 1/4. This gave birth to the Bekenstein-Hawking entropy-area (BHAE) law \cite{bekenstein,hawking}, $S = \frac{A}{4}$. For systems which are thermalized by the acceleration, the associated Rindler horizon also obeys the same area-entropy law due to presence of the FDU temperature \cite{satz}. As such, for highly accelerated systems, one can directly confirm the hypothesis of Bekenstein and Hawking.

The change in the horizon area is determined by the amount of energy radiated by each positron, $\Delta \tilde{E}$, into the horizon. This quantity we will extract from the data via $\Delta \tilde{E}=\int\frac{d\mathcal{S}_{data}}{d\omega}d\omega dt$. The corresponding change in horizon area \cite{satz} is given by $\Delta A = \frac{G c^{5}}{\hbar} \frac{8 \pi m^{3} \Delta \tilde{E}}{E_{i}^{3} a }$. Here, $E_{i}$, is the initial beam energy minus the cutoff frequency. By integrating the first law of thermodynamics, using our radiation reaction temperature, Eq. (4), we have and entropy difference, $\Delta S = \frac{c^{8}}{\hbar^{2}} \frac{\pi m^3  }{ a} \lbk \frac{1}{(E_{i} - \Delta \tilde{E})^{2}}-\frac{1}{E^{2}_{i}} \rbk$. As such, our area to entropy ratio is given by \cite{aqd}, 
\bqa
\frac{\Delta A}{\Delta S} = \ell^{2}_{p}\frac{8 \Delta \tilde{E}}{E^{3}_{i}}\lbk \frac{1}{(E_{i} - \Delta\tilde{E})^{2}} -\frac{1}{E^{2}_{i}} \rbk^{-1}. \label{bh}
\eqa 

The convergence of this expression for all $\Delta \tilde{E}$ and $E_{i}$ throughout the integration of the data set to $4 \ell_{p}^{2}$ will not only imply thermality within the data, but also confirm the BHAE law. The integration over the data is presented in Fig. 3 below.

\begin{figure}[H]
\centering  
\includegraphics[scale=.29]{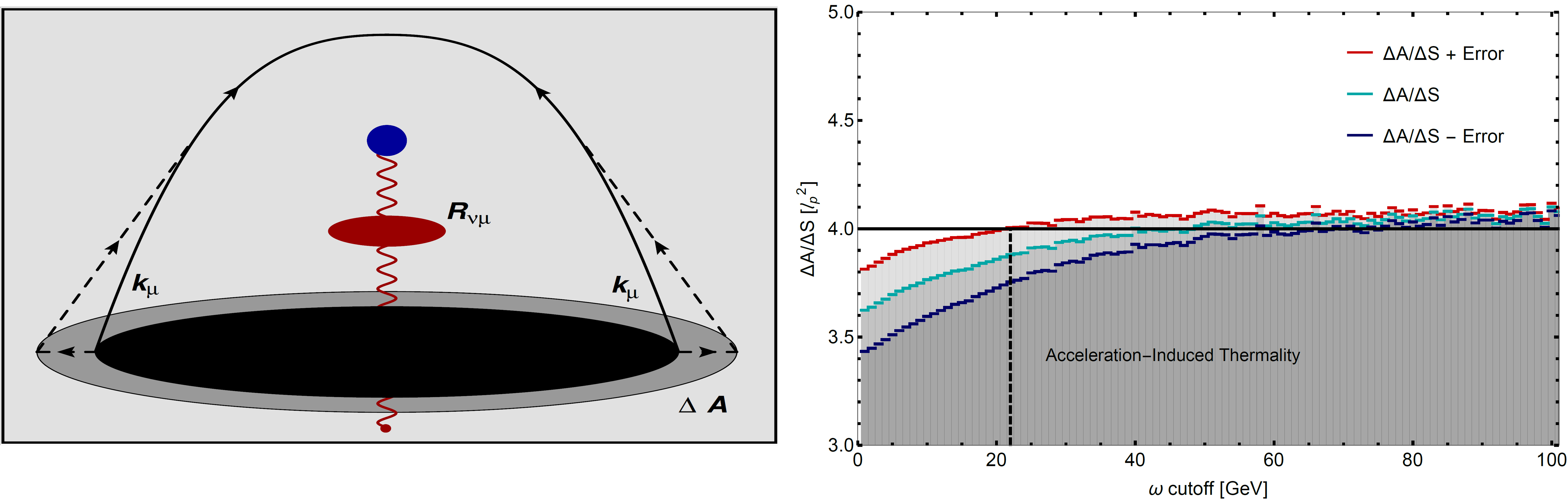} 
  \caption{\textbf{Rindler horizon area change and integration of area-entropy ratio:} (Left) demonstrates the area change of the Rindler horizon based on the geodesic deviation of null rays which characterize the horizon via the curvature sourced by the energy-momentum of the photons emitted. (Right) shows the integration of the area-entropy ratio across the data as a function of low energy cutoff. The convergence to $4 \ell_{p}^{2}$ occurs at the same low energy threshold of $\sim 22$ GeV as the thermalization time.}
\end{figure} 

 These experimental measurements confirm the predictions of Bekenstein \cite{bekenstein} and Hawking \cite{hawking} in regards to the relationship between horizon area and entropy along with its extention to Rindler horizons by Bianchi and Satz \cite{satz}. 

\section{Conclusions}

In this essay, we presented notes on the recent experimental observation of acceleration-induced thermality \cite{aqd}. In short, a high energy channeling radiation experiment carried out by CERN-NA63 \cite{wistisen} was successful in measuring radiation reaction, the Unruh effect, and the Bekenstein-Hawking area-entropy law.

\section{Acknowledgments}
We thank Ulrik Uggerhoj, George Matsas, and Niayesh Afshordi for their support.

\goodbreak

\pagebreak

\end{document}